\documentstyle[prl,aps,twocolumn,epsf]{revtex}
\begin{document}
\draft

\twocolumn[\hsize\textwidth\columnwidth\hsize\csname
@twocolumnfalse\endcsname
\title{Search for Multiple Step Integer Quantum Hall Transitions}
\author{Xin Wan and R. N. Bhatt}
\address{Department of Electrical Engineering, 
Princeton University, 
Princeton, NJ 08544}
\date{\today}
\maketitle 

\begin{abstract}
Recent experiments in the integer quantum Hall
regime seem to find
{\it direct} transitions from a quantum Hall
state with Hall conductance $\sigma _{xy} = n e^2/h $
with integer $n > 1$, to an insulating state in weak magnetic fields. 
We study this issue using a variation 
of the tight-binding lattice model for
non-interacting electrons.
Although quantum Hall transitions with change in Hall conductance
$ne^2/ h$ with $n > 1$ do exist in our model for special tuning
of parameters, they generically split into quantum Hall 
transitions with the Hall
conductance changing by $e^2 / h$ at each transition.
This suggests that a generic multiple step quantum Hall transition
is incompatible with a non-interacting electron picture.
\end{abstract}

]

For non-interacting two-dimensional electron systems
in the presence of a perpendicular applied
magnetic field ($B$), Kivelson, Lee and Zhang~\cite{kivelson92}
proposed a global phase diagram, which predicted that in
a spin-polarized system, at a quantum Hall transition, the Hall
conductance
($\sigma_{xy}$) can only change by $e^{2} / h$ as filling
factor ($n$)
changes in the integer quantum Hall effect (IQHE) regime. 
Stimulated by Kivelson {\it et al.}'s proposal, 
a number of experiments reported
the observations of $\sigma_{xy} = 0 \rightarrow 2 e^{2} / h
\rightarrow 0$  
transition \cite{jiang93,hughes94,wang94}, which is consistent
with the global phase diagram, 
assuming the Zeeman splitting in the lowest Landau
level is unresolved.  However, more recently, 
many experiments~\cite{shahar95,kravchenko95,song97,hilke00},
some in systems with spin-split Landau bands,
observed direct transitions from an insulating state at
low fields to $ \sigma_{xy} = n e^2 /h$ IQHE
states with $n$ upto 6. Henceforth in this letter, we
dub these transitions from the insulating phase to $n > 1$ IQHE
states {\it multiple step quantum Hall transitions} (MSQHTs).

In a series of papers~\cite{sheng97,sheng98,sheng99,sheng00}, 
Sheng {\it et al.} claimed that the experimental studies of the
quantum Hall transitions in weak magnetic fields are consistent
with their numerical calculations on a tight-binding lattice
model. 
In particular, the MSQHTs are the consequence of the critical
energy in the lowest Landau level floating and merging 
with critical energies in higher Landau levels~\cite{sheng00}.
However, it remains unclear how these
critical energies can merge, and thus give rise to an MSQHT.
An alternative interpretation of the apparent MSQHT is a
series of single step ($\Delta \sigma_{xy} = e^2/h$) 
plateau transitions that are
too close to distinguish numerically.  

The two scenarios may be resolved by studing the critical behavior
in the vicinity of
the transition(s) under investigation. 
For a single, spin-split Landau level, numerical
studies by various groups~\cite{huo92,huckestein95,liu94,chalker88}
using different microscopic Hamiltonians
are consistent with the same critical behavior at the quantum Hall
transition (QHT). 
In all cases, the localization length 
diverges with an exponent $\nu \simeq 2.3 \pm 0.1$
at the critical energy, $E_c$, which is at the band center in the case
of
electron-hole symmetry.
By studying the localization length exponent, for instance, 
Lee and Chalker~\cite{lee94} demonstrated, 
in a spin-degenerate Landau level, the existence of a
pair of delocalization transitions in the same universality class 
as the spin-split system.

In this letter, we report our search for MSQHTs by truncating the
Hilbert space of the tight-binding lattice model to a single (central)
magnetic subband, 
which contains a multiple of $2 e^2 /h$ Hall conductance,
and can therefore support a MSQHT in principle.
While this model does not correspond directly to the
experimental situation, it allows for a much clearer
interpretation regarding the possible existence of
a multiple step transition.
We also expect the study to lead to a clearer understanding of the
tight-binding lattice in the context of weak magnetic fields. 

Our results can be summarized as follows:
we find that the system can support a MSQHT
in the limit of infinite disorder, but in this limit
the multiple step comes from a set of noninteracting, single
step transitions with accidental degeneracy of the step
position. As soon as the disorder is made finite, the mutiple
step breaks into single steps, whose critical behavior is the
same as that for isolated Landau levels.  

We consider a tight-binding model of non-interacting spinless electrons
on two-dimensional square lattice,  
with nearest neighbor hopping ($t$), a uniform perpendicular magnetic
field
$B$, and on-site disorder. 
The Hamiltonian can be written as:
\begin{equation}
H = - t \sum_{\langle ij \rangle}{(e^{i 2 \pi a_{ij}}
c^{\dagger}_{i} c_{j}
        + h.c.)} + \sum_{i}{\epsilon_{i} c^{\dagger}_{i}
c_{i}}, 
\end{equation}
where $c^{\dagger}_{i}$ ($c_{i}$) is fermionic creation (annihilation)
operators on site $i$,
$\langle ij \rangle$ are nearest neighbor sites, 
$t$ is the hopping strength, and $a_{ij}$ are phase factors
due to the magnetic field $B$.
$\epsilon_{i}$ is a random on-site potential, with a
distribution
$ P(\epsilon) = ( 2 \pi \sigma^2)^{-1/2}
        \exp \left ( - {\epsilon^2} / {2 \sigma^2} \right ) $.
The magnetic flux per unit cell is chosen to be $\phi / \phi_0
=
\sum_{\Box}{a_{ij}} = {1 / (2N+1) }$ with integer $N$, where
$\phi_0 =
hc / e$ is the flux quantum. 

In the presence of $B$, the tight-binding energy spectrum splits into
($2N + 1$) magnetic subbands. 
Each of the $2N$ side subbands carries a Hall conductance $\sigma_{xy} =
e^2 / h$,
while the
center subband carries $\sigma_{xy} = - 2N e^2 / h$. 
In contrast to other
approaches\cite{sheng97,sheng98,sheng99,sheng00,yang96},
we {\em first truncate} the Hamiltonian to the Hilbert
space spanned by the eigenstates in the center subband in the
absence of disorder, before adding the disorder.  
This results in a single subband with $\sigma_{xy}= - 2N
e^2 / h$ for any disorder strength.
The negative sign of $\sigma_{xy}$ is not essential to the
discussion
here, since we are primarily interested in the scaling behavior
near the critical point(s) of the
center subband.
 
We calculate, within the center subband, the Thouless
number~\cite{thouless}, 
as a measure of the diagonal 
conductance $ \sigma_{xx} $ in units of $ e^2/h$,
for different lattice sizes $ L \times L $ ($L$ is
the number of sites in each dimension):  
\begin{equation}
g_L (E) = {\langle | \delta E | \rangle \over \Delta E}
\sim {h \over e^2} \sigma_{xx}, 
\end{equation}
where $\langle | \delta E | \rangle$ is the average shift in
the energy
level due to a change of boundary condition from periodic to
anti-periodic, and $\Delta E = 1 / L^2 \rho(E)$ 
the mean energy level separation, $ \rho(E) $ being the density of
states.  
$g_L (E)$ is dimensionless, and by finite-size scaling,
is expected to be a function of $L$ only through the ratio $ L/ \xi $,
where $ \xi $ is the localization length characterizing the
wavefunctions
at the energy E, which diverges at critical energies $E_c$ as
$ \xi \sim | E - E_c | ^ { - \nu } $. As a consequence, the square root
of the second moment of
$ g_L (E) $, 
\begin{equation}
\label{eq:TNWidth}
W(L) = \left [ { \int_{-\infty}^{\infty} (E - E_c)^2 g_L (E) dE
        \over \int_{-\infty}^{\infty} g_L (E) dE} \right ]^{1/2} .
\end{equation}
is expected to scale as
$W(L) \sim L^{-1 / \nu}$ in the thermodynamic limit for a single
critical energy.  

\begin{figure}[hbt]
\epsfxsize=3.2in
\centerline{\epsffile{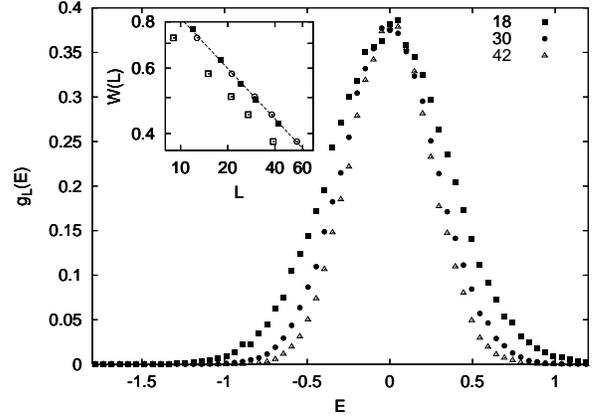}}
\caption{ 
\label{infinite3}
Thouless number $g_L(E)$ of the center subband ($N = 1$) with
infinite
disorder ($\sigma = 1$ and $t = 0$) for lattices with size $L$ =
18, 30 and 42. 
The inset shows the width $W(L)$, defined by Eq.~(\ref{eq:TNWidth}), of
corresponding $g_L(E)$, on log-log scale. 
Note the difference between even $L$ (solid squares) 
and odd $L$ (open squares); 
as explained in the text, when shifted by multiplying $L$ by a factor of
$\sqrt{2}$, results for odd-sized lattices (open circles) fall onto the
dashed line fitted to even-sized lattices (solid squares).  
}
\end{figure}

We first consider the case $\phi / \phi_0 =
1 / 3$, where the central subband carries $\sigma_{xy} =  -2e^2 / h$, 
and to further simplify the situation, we begin with the infinite
disorder
limit ($\sigma / t \rightarrow \infty$). 
Equivalently, we set $t = 0$ {\em after} truncating the Hilbert
space to
the center subband only, so that the energy scale is finite.
Figure~\ref{infinite3} shows $g_L(E)$ for different (even) L,
averaged over 250 to 2500 samples, depending on $L$.
The inset shows the width $ W(L) $ of the curves {\em vs.} $L$
(solid squares for even $L$) on a double logarithmic plot.
The straight line fit implies a {\em single} critical
energy $ E_c = 0 $, and a critical exponent $\nu = 2.2 \pm 0.1$,
consistent with the value obtained for a {\em single} step
quantum Hall transition in a sideband \cite{wan00a} or an isolated
Landau level\cite{huo92}. 
The data are well fit by a gaussian form,
$ g_L(E) = g_0 exp (- E^2 / 2 [W(L)]^2 ) $, with $ g_0 \simeq 0.38 $,
approximately {\em twice} the value
$g_0 \simeq
0.2$
in a side subband\cite{wan00a,yang00}.

To understand this result, we examined the structure of the
projected Hamiltonian. Upon application of the projection operator,
the on-site random potential becomes non-local, giving rise to
(short-range) hopping. 
In the $ t = 0 $ limit, there is no direct hopping, and
the hopping
generated by the random on-site potential due to projection,
leads to a bipartite Hamiltonian.  
Thus, for even-sized lattices, the system splits into
two {\it disconnected} sublattices interpenetrating each other,
with {\it identical} statistical properties, since hopping from one 
sublattice to the other is not generated.
This explaines why we see a {\em single} QHT
at $E=0 $, with the same value of $ \nu $ as a single-step QHT, 
but {\em twice} the Thouless conductance $ g_0 $ - {\it we simply have
two independent transitions at the same energy}.

The above result is confirmed by the observation that for {\it odd} $L$,
where hopping from one sublattice to the other is allowed
by the boundary conditions, the width $W(L)$ are quite different
from those for even $L$ (see solid squares in the inset of
Fig.~\ref{infinite3}. 
This seemingly strange result can be understood by 
rearranging one sublattice by translating its members appropriately by
$L$
lattice
constants along one of the sides of the original lattice.
One thus
``unfolds" an odd-sized lattice to a new square lattice rotated
by $\pi
/ 4$ and enlarged by $\sqrt{2}$ in linear size. 
By replotting $W(L)$ of odd $L$ as a function of
$\sqrt{2}L$,
which is equivalent to translating the original points to the
right on 
the log-log plot (Fig.~\ref{infinite3} inset), we found that
these points fall
onto the {\em same} straight line obtained from those of even-sized
lattices. 
Therefore, the Hamiltonian in an odd-sized lattice describes a
single
quantum Hall transition on a ``folded'' lattice.
Study of statistical properties of energy spectrum have also
confirmed that
the odd-even anomaly is the consequence of whether the
Hamiltonian is bipartite~\cite{wan00a}. 

\begin{figure}[hbt]
\epsfxsize=3.2in
\centerline{\epsffile{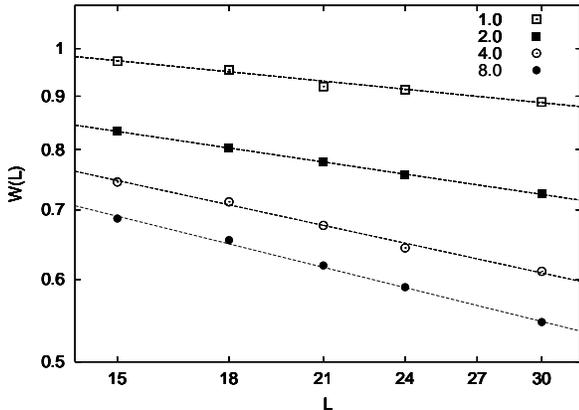}}
\caption{ 
\label{finite3}
Width of Thouless number curves, $W(L)$,
as a function of lattice size $L$ for $\sigma / t =
1$, 2, 4, and 8. }
\end{figure}

We next consider the more generic case, with finite
disorder. If, in the thermodynamic limit, there remains
a single critical energy at the band center, one would expect
curves for $W(L)$ versus $L$ to be straight lines on a log-log
plot for large sizes, with a universal slope, 
related to the localization length critical
exponent for a double step QHT. Instead, we find (see
Fig.~\ref{finite3})
a series of straight lines with a slope that depends on the
disorder ($ \sigma /t $); furthermore, the peak value, 
$g_L(0)$ is dependent on $ \sigma /t $ as well.
(Note that the odd-even effect gets smaller as $\sigma /t$ decreases).
Non-universal $\nu$ for other disorder distributions $ P(\epsilon) $
have been observed as well~\cite{wan00a}.   

\begin{figure}[hbt]
\epsfxsize=3.2in
\centerline{\epsffile{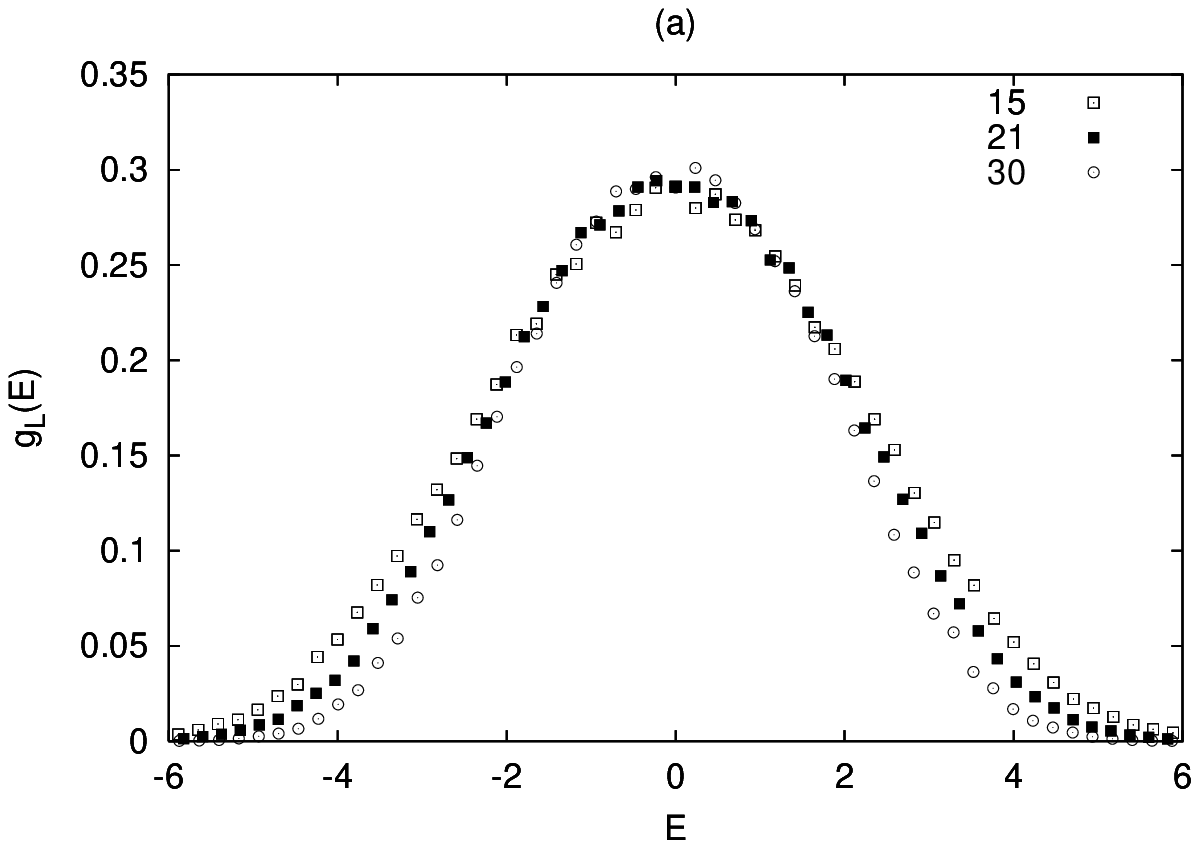}}
\epsfxsize=3.2in
\centerline{\epsffile{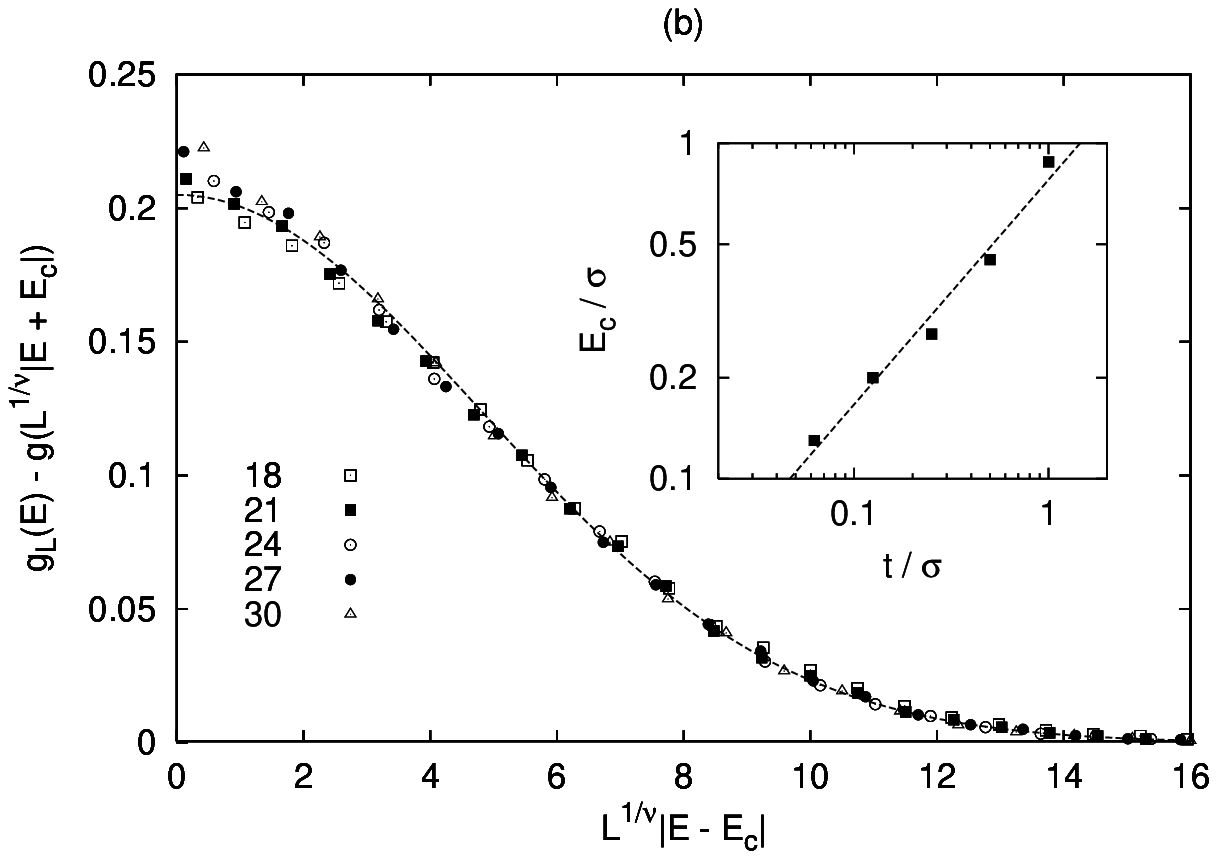}}
\caption{ 
\label{scaling}
(a) Thouless number $g_L(E)$ of the center subband ($N = 1$) of
lattices of linear dimensions $L = 15$, 21 and 30 
for $\sigma / t = 5$. 
(b) Scaling curve of the Thouless number, $g(L^{1 / \nu} |E -
E_c|) $, 
with $E > E_c = 1.3$ and $\nu = 2.5$. 
Dashed line shows the gaussian curve found for single step quantum Hall
transitions. 
Inset shows that the $E_c$, in units of $\sigma$,
can be
fit to a power law (dashed line) as a function of $t
/ \sigma$. 
} 
\end{figure}

Figure~\ref{scaling}(a) shows the full Thouless number curves
for $ \sigma /t = 5 $ for $L = 15, 21, 30$, averaged over
120-600 samples; the curves show
a distinct departure from the gaussian form of the scaling curve
for the single-step QHT, with a pronounced flat-top shape.

The apparent non-universal exponent is based on the
assumption that the critical energy remains at $ E = 0 $;
if instead, there were two single step QHT at
$\pm E_c$ in the thermodynamic limit, one would
expect the $g_L(E)$ curves to be made up of two
contributions. Indeed, a fit of the form
\begin{equation}
\label{de}
g_L (E) = g ( L^{1 / \nu} |E - E_c|) + g ( L^{1 / \nu} |E +
E_c|),
\end{equation}
is found to work extremely well for all $|E| \ge E_c$
for {\em all} values of the disorder studied 
with a disorder dependent critical energy $E_c$ with a single value of
$\nu$.
The universal exponent obtained from such a fit is 
$\nu = 2.3 \pm 0.2$, which
is just the result for a single step QHT. Furthermore,
the shape of the scaling curve $ g(L^{1/\nu} |E - E_c|) $
(see Fig.~\ref{scaling}(b)), {\it is very well fit the gaussian form
obtained for a single step QHT}. 
Finally, the dependence of $E_c$ on $\sigma$,
shown in the inset of Fig.~\ref{scaling}(b), is found to be
$E_c / \sigma = 0.78 \left ( t / \sigma \right
)^{0.67}$ for $ 0.03 < \sigma / t < 1.0 $,
consistent with $E_c / \sigma \rightarrow 0$ in the 
limit of $t / \sigma \rightarrow 0$. 
The fit of Eq.~(\ref{de}) does not
work as well {\it between} the two critical energies
presumably because of the extremely large localization lengths,
much beyond the sizes of systems studied, for this region. 
Thus the data implies
that the accidental double step for infinite disorder splits
into two single steps for the generic, finite disorder case,
leading to a behavior similar to that obtained in double-layer
systems\cite{sorensen96} or for electrons with spin-orbit
coupling~\cite{lee94}.

\begin{figure}[hbt]
\epsfxsize=3.2in
\centerline{\epsffile{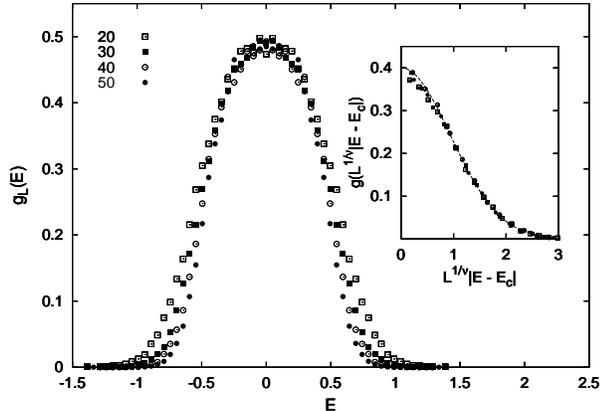}}
\caption{ 
\label{infinite5}
Thouless number $g_L(E)$ of the center subband ($N = 2$) with
infinite
disorder ($\sigma = 1$, $t = 0$) for lattices with $L = 20$,
30, 40,
and 50.  
The inset shows the scaling collapse of the Thouless
number
$g(L^{1 / \nu} |E - E_c|)$, using 
the four $g(E)$ curves for $E > E_c = 0.3$ and $\nu = 2.4$.
} 
\end{figure}

We next consider the case with $\phi / \phi_0 = 1/5$ ($N = 2$), in
which the center subband has total Hall conductance $-4 e^2 / h$.
Since this case allows for many more possibilities, we restrict our
attention to the infinite disorder ($ t=0 $) case. In this limit
we again obtain a Hamiltonian that is bipartite for even-sized lattices.
This implies that the transition in the center subband can be
either (a)
two double step quantum Hall transitions at $E=0$, or (b) a pair of
split,
single step
quantum Hall transitions, each pair occurring at $ E = \pm E_c $.
Figure~\ref{infinite5} shows $g_L(E)$ of even-size lattices for
infinite
disorder, averaged over 80 to 1800 samples, depending
on system size. 

Forcing the width $ W(L) $ to scale as a
power law in $L$ over the sizes studied yields an exponent 
$ \nu = 4.2 $. However, the
shape of the $ g_L (E) $ curves is
rather flat-topped compared to a gaussian, and the almost doubled
value of $ \nu $ also suggests that a more natural scenario is
(b) above. Indeed, fitting to the form given by Eq.~(\ref{de}) for
$ |E| > E_c $ yields
$E_c = 0.3$, $g_0 = 0.4$, $\nu = 2.4$, and {\em furthermore}
a gaussian shape for the scaling curve (see inset of Fig 4).
Thus, with a single adjustable parameter, $E_c$, the data are consistent
with a pair of degenerate single step transitions, with exactly
the same Thouless number peak, critical exponent, and scaling curve.
This splitting of the critical energy in each sublattice argues strongly
for single step quantum Hall transitions being the generic situation for
non-interacting electrons. 

Our study has raised questions about the interpretation of the results
found in the full
tight-binding lattice model~\cite{sheng97,sheng98,sheng99,sheng00}.
In particular, a genuine MSQHT must be associated with degenerate
critical energies, which is less likely to appear in the full model with
finite amount of disorder and critical energies in all side subbands
that can interact with those in the center subband. 
Very recently, Koschny {\it et al.}~\cite{koschny01} have claimed to
resolve
two separate transitions with filling factor $n = 2$ in the full lattice
model, and thus contradicted the direct transition from insulator to
high
Hall plateaus suggested by Sheng {\it et al.}
Nevertheless, the difficulty of unambiguously resolving this issue,
as well as the obstacle that forbids us to obtain the scaling behavior
between adjacent critical energies, is consistent with the existence of 
the large localization length well beyond the sizes accessible to
present numerical studies.  
The length scale may, indeed, agree with the large crossover
length
argued by Huckestein~\cite{huckestein00} in the framework of the
standard scaling theory of the IQHE. 

To summarize, we have found that for non-interacting electrons,
even in a single band with total Hall conductance
a multiple of $2 e^2 / h$, in the presence of disorder, a
sequence of single step quantum Hall transitions 
appears to be the generic situation. 
The only multiple step transition we have been able to obtain
is by 
special tuning of parameters, for which there exist accidental 
degeneracies of critical energies along with zero coupling of
these 
degenerate critical energies.
In such cases, the multiple step transition is a trivial
superposition of 
single step transitions, with the same exponents for the
localization length 
($\nu = 2.3$) and scaling curves for the Thouless conductance.
This calls into question the interpretation of the apparent
multiple step transitions seen in 
experiment~\cite{shahar95,kravchenko95,song97,hilke00}
using a non-interacting model\cite{sheng97,sheng98,sheng99,sheng00}.

This work was supported by NSF DMR-9809483. We thank Kun Yang for
discussions. 


\begin{thebibliography}{10}

\bibitem{kivelson92}
S. Kivelson, D.~H. Lee, and S.~C. Zhang, Phys. Rev. B {\bf 46},  2223 
(1992).

\bibitem{jiang93}
H.~W. Jiang, C.~E. Johnson, K.~L. Wang, and S.~T. Hannahs, Phys. Rev.
Lett.
  {\bf 71},  1439  (1993).

\bibitem{hughes94}
R.~J.~F. Hughes {\it et~al.}, J. Phys. Condens. Matter {\bf 6},  4763 
(1994).

\bibitem{wang94}
T. Wang {\it et~al.}, Phys. Rev. Lett. {\bf 72},  709  (1994).

\bibitem{shahar95}
D. Shahar, D.~C. Tsui, and J.~E. Cunningham, Phys. Rev. B {\bf 52}, 
R14372
  (1995).

\bibitem{kravchenko95}
S.~V. Kravchenko, W. Mason, J.~E. Furneaux, and V.~M. Pudalov, Phys.
Rev. Lett.
  {\bf 75},  910  (1995).

\bibitem{song97}
S.~H. Song {\it et~al.}, Phys. Rev. Lett. {\bf 78},  2200  (1997).

\bibitem{hilke00}
M. Hilke {\it et~al.}, Phys. Rev. B {\bf 62},  6940  (2000).

\bibitem{sheng97}
D.~N. Sheng and Z.~Y. Weng, Phys. Rev. Lett. {\bf 78},  318  (1997).

\bibitem{sheng98}
D.~N. Sheng and Z.~Y. Weng, Phys. Rev. Lett. {\bf 80},  580  (1998).

\bibitem{sheng99}
D.~N. Sheng and Z.~Y. Weng, cond-mat/9906261.

\bibitem{sheng00}
D.~N. Sheng, Z.~Y. Weng, and X.~G. Wen, cond-mat/0003117.

\bibitem{huo92}
Y. Huo and R.~N. Bhatt, Phys. Rev. Lett. {\bf 68},  1375  (1992).

\bibitem{huckestein95} B. Huckestein, 
Rev. Mod. Phys. {\bf 67}, 357 (1995).

\bibitem{liu94} D. Liu and S. Das Sarma, 
Phys. Rev. B {\bf 49}, 2677 (1994).

\bibitem{chalker88} J.~T. Chalker and P.~D. Coddington, 
J. Phys. C {\bf 21}, 2665 (1988).

\bibitem{lee94} D.~K.~K. Lee and J.~T. Chalker, 
Phys. Rev. Lett. {\bf 72}, 1510 (1994).

\bibitem{yang96}
K. Yang and R.~N. Bhatt, Phys. Rev. Lett. {\bf 76},  1316  (1996).

\bibitem{thouless} J.~T. Edwards and D.~J. Thouless, 
J. Phys. C {\bf 5}, 807 (1972);
D.~J. Thouless, Phys. Rep. {\bf 13}, 93 (1974);
D.~C. Licciardello and D.~J. Thouless, J. Phys. C {\bf 8},  4157 
(1975).

\bibitem{wan00a}
X. Wan, Ph.D. thesis, Princeton University, 2000.

\bibitem{yang00}
K. Yang {\it et~al.}, J. Phys. Condens. Matter {\bf 12},  5343  (2000).

\bibitem{sorensen96}
E.~S. S{\o}rensen and A.~H. MacDonald, Phys. Rev. B {\bf 54},  10675 
(1996).

\bibitem{koschny01} Th. Koschny, H. Potempa, and L. Schweitzer, 
cond-mat/0102394. 

\bibitem{huckestein00} B. Huckestein, 
Phys. Rev. Lett.  {\bf 84},  3141  (2000). 

\end{thebibliography}

\end{document}